\documentclass[aps,pre,twocolumn,groupedaddress,amssymb,floatfix,showkeys,showpacs]{revtex4-1}

\usepackage{graphicx, graphics, epstopdf, algorithm, color,datetime}
\newtheorem{definition}{Definition}

\begin{document}


\title{Ensemble approaches for improving community detection methods}


\author{Johan Dahlin}
\email[]{johan.dahlin@gmail.com}
\thanks{Current address: Department of Electrical Engineering, Division of Automatic Control, Linkoping University}
\affiliation{Department of Physics, Ume{\aa} University}
\affiliation{Division of Information and Aeronautics Systems, Swedish Defence Research Agency}

\author{Pontus Svenson}
\email[]{ponsve@foi.se}
\affiliation{Division of Information and Aeronautics Systems, Swedish Defence Research Agency}


\date{Sep 1, 2013}

\begin{abstract}
Statistical estimates can often be improved by fusion of data from several different sources.
One example is so-called ensemble methods which have been successfully applied in areas such as machine learning for classification and clustering. 
In this paper, we present an ensemble method to improve community detection by 
aggregating the information found in an ensemble of community structures. 
This ensemble can found by re-sampling methods, multiple runs of a stochastic community detection method, or by several different community detection algorithms applied to the same network. The proposed method is evaluated using random networks with community structures and compared with two commonly used community detection methods. The proposed method when applied on a stochastic community detection algorithm performs well with low computational complexity, thus offering both a new approach to community detection and an additional community detection method.
\end{abstract}


\maketitle


\section{Introduction}
Networks 
are ubiquitous in nature and provide versatile models for many-body systems with non-regular interactions. For these reasons, they have become an important topic of current research.  Network science has
provided novel application areas for methods from statistical physics, and has in turn developed new methods that can be used to study physical systems.
The study of networks is often concerned with quantifying different microscopic aspects of the structure, such as centrality measures, degree distributions, information flow in networks, and robustness.  Concepts and methods from network analysis have been applied to a large range of different types of networks. Some examples of the most important applications include analysis of energy grids, epidemiology, metabolic networks, protein-protein interaction networks, social networks, etc. \cite{clustering-santo} . 
Networks often arise as a consequence of strongly interacting agents or many-body systems
with weak interactions but  where the complex network structure leads to emergent behavior that has interesting physical properties.

Other important topics in network research include higher-level structures called communities.
Communities are commonly defined as  groups of nodes more densely connected with each other than with nodes outside the group. These communities, or partitions, are commonly found in e.g.\ our everyday social networks as colleagues, high-school friends, neighbors, family, etc. Much effort has been devoted to defining community structures and finding algorithms to detect partitions with low computational complexity and high accuracy. Despite this, no accepted general definition of what a community structure is has been proposed. In order to clarify the proper (if any) definition of community structures, there is a need for new approaches. In this paper, we present a new approach to the community detection problem by considering an {\em ensemble} of community clustering methods working on the same network. The results of the different community detection algorithms are then fused into a, hopefully more accurate, representation of the community structure of the network.

It is our hope that the work presented here will contribute both to more efficient algorithms for community detection and to a conceptual discussion about what a community structure is. By considering an ensemble of clustering methods, it is possible to consider different definitions of community structure. More effective algorithms can be found by merging (aggregating) many runs of fast stochastic algorithms as well as several runs of the same algorithm using different settings. 
In addition  the latter method can be used to analyze the community structure of the network at many different scales, providing insight into the relations between community structure at different levels.
The merging method is also applicable in aggregation of communities generated by bootstrap replicates of the network data, which is necessary in
cases where there is missing or incomplete information available about the network.

Previous and related work includes work in the areas of community detection and in ensemble methods developed for data clustering and classification methods. 
In recent years a large number of methods for detecting community structures have been developed, drawing on knowledge from many different fields, e.g.\ statistical mechanics, discrete mathematics, computer science, statistics, and sociology. These methods have also been improved to handle weighted, directed, and multi graphs. A thorough review of the current state in community detection is given in 
Ref.\cite{clustering-santo}; we provide some 
background in Section \ref{sec:community}.

Ensemble clustering is a technique used in e.g.\ bioinformatics applications and is useful in merging several clustering results into one. To our knowledge, no work has been devoted to applying these methods in community detection problems. However other methods have been used to merge several community structures, e.g.\ voting in Refs.\ \cite{citeulike:1724653,citeulike:6597911}. As data clustering and community detection are quite similar, it should be possible to merge communities in the same manner as ensembles of data with good results. Ensemble clustering methods were first introduced in Ref.\ \cite{strehl03} and further developed by e.g.\ Refs.\ \cite{citeulike:1267551,citeulike:1319014}. 

This paper continues with a presentation of the background in community detection (Section~\ref{sec:community}) and ensemble methods (Section~\ref{sec:ensemble}) previously used in classification and clustering, where we
discuss some common community detection methods and the performance of ensemble methods 
The ensemble-based community clustering method is introduced in Section~\ref{sec:algorithm}, where its computational complexity is discussed and suggestions for 
how to estimate the certainty of the obtained solution given.
 Finally, Section~\ref{sec:experiments} offers some simulation experiments which compare the proposed method to two well-known community detection algorithms: greedy modularity maximizing agglomerative algorithm \cite{citeulike:2390590}, and a q-Potts based spin glass model \cite{Reichardt}.

 The paper is concluded with a summary with some remarks concerning implications and future work.

\section{Community detection}
\label{sec:community}
Networks consist of nodes, representing e.g.\ individuals, computers, or proteins, that are connected by edges representing e.g.\ friendships, network connections, or other types of interactions. Formally, networks are defined using graph structures, $G=(V,E)$, where $V$ denotes the set of nodes and $E$ the set of edges. We denote the number of edges in a graph  $n=|V|$ and the number of edges by $m=|E|$.

Networks often contain some form of community structure, i.e.\ groups of nodes that are more densely connected to each other than to nodes outside of the group. In essence, this resembles the similar problem in data clustering, where similar data points are grouped together into clusters. In the same manner, nodes inside communities are often thought of as sharing some common feature. The interpretation of this feature naturally depends on the nature of the network data, e.g.\ communities in social networks are often thought of as constituting some social group sharing family ties, employer, a specific interest, etc. Detection of communities is therefore an important tool in sociology and other related areas, but is also used in fields including ecology and biology where food webs, protein-protein interaction, metabolic networks and natural resource exploittation networks are of interest\cite{bodin}.

Community detection is a widely studied subject and much work has been devoted to developing faster and more accurate automatic methods for detection and verification of communities in complex networks. This section serves only as a short review of the field and some of the proposed methods for identifying communities in networks. For a comprehensive review of the field as a whole, we refer the reader to Refs.\ \cite{clustering-santo} and \cite{citeulike:4091170}.

\subsection{Existence and uniqueness}
There is no formal generally accepted
definition of what a community, despite large efforts in the study of community detection and complex networks. In this paper, we adopt the practical viewpoint of \textit{Definition \ref{def:community}} and use the definition due to Ref.\ \cite{Radicchi} for what a community structure is.

\begin{definition}[Community]
A community (in qualitative terms) is a subset of nodes within a network such that connections between nodes in the subset are denser than connections with the rest of the network.
\label{def:community}
\end{definition}

The main problem with this definition is questions like: how large a subset must be (can a community consist of only a few nodes?) and what exactly denser means in terms of number of edges inside the community versus between communities.
We return to the latter question in connection with the discussion of algorithms for generating synthetic (random) networks with community structures in below.
Another issue with the definition is that in real networks, there are often edges of different kinds. For example, a social network contains edges that denote friendship, which are separate from those that represent colleagues. When
looking for work-related communities, only work-related edge types should be considered.

\begin{figure}[h]
	\centering
	\includegraphics[width=0.30\textwidth,angle=0]{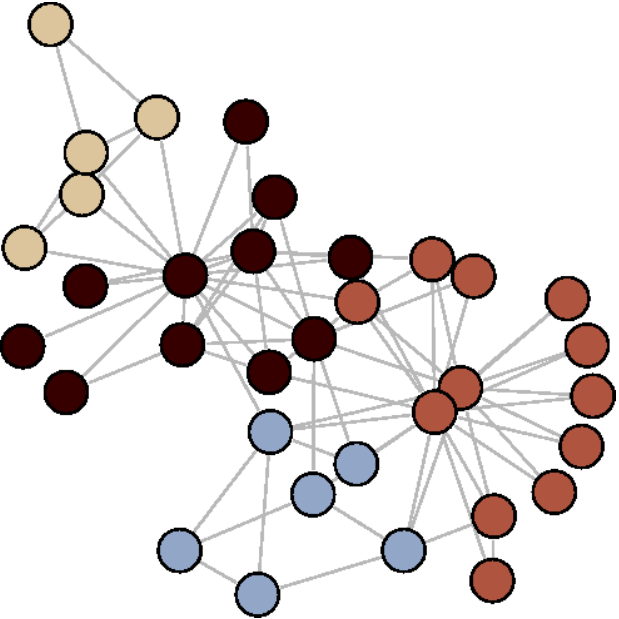}
	\caption{\small{Four communities found using the q-Potts spin glass method (see below) in the famous karate network from Ref.\ \cite{zachary1977ifm}, indicating the friendships in a karate club at an U.S. university during the 1970's.}}
    	\label{fig:karatecommunities}
\end{figure}

The lack of a general definition  raises problems related to uniqueness and existence of communities in networks. A network may contain many different community structures depending on the scale considered, from just one community containing all nodes to communities only containing one node each. This is known from previous work as the \emph{resolution limit}, when discussing modularity (see next subsection) as a quality measure for community structures. Problems with existence includes questions regarding if the communities are the result of the data or by the underlying process generating this data. With this it is meant that observations seldom identifies the entire network and therefore it is difficult to verify if the resulting communities exist or only is an artifact of some missing or erroneous data.

Assume that we have studied a dense social network and have identified some of the ties between varies individuals. As previously discussed, it is often difficult to identify all ties in the network and therefore we may only find a subset of all the edges in the real network. Applying standard community detection methods on this network will probably return some community structure with a certain number of communities. But as only a subset of the edges have been used, the real unobserved network may only contain one community. Therefore, the existence of the identified communities are in question, do they really exist or not?

The previous example discusses the well-known problem of the \emph{robustness} of communities. This has been studied by e.g.\ using bootstrap methods to generate subsets of edges and study how the community structure appear as a mean of a large number of bootstrapped networks. Later, we will see how the  methods proposed in this paper offer another solution to the problem of estimating the robustness of a community structure.

Additional problems with the definition of community structure arises because of the multi-modal nature of most networks of interest. If we for instance are interested in clustering a social network of individuals and their relations, we must distinguish between many different kinds of interpersonal relations: friendship, colleague, co-author and citation networks. 

\subsection{Quality of community structures}
Comparing the quality of a obtained community structure is usually done by a measure called \emph{modularity}. This measure compares the network structure with the structure of a null model in which edges are randomly redistributed keeping the degree of all nodes fixed. The concept of modularity was introduced in Ref.\ \cite{citeulike:686555} and is usually denoted $Q(\mathbf{c})$, where $\mathbf{c}=(c_i)$ is the vector of communities to which each node $i$ belongs. The measure is calculated using the  expression

\begin{eqnarray}
Q(\mathbf{c}) &=& \frac{1}{2m} \sum_{i,j=1}^n \left( A_{ij} - N_{ij} \right) \delta(c_i,c_j) \nonumber \\ &=& \frac{1}{2m} \sum_{i,j=1}^n \left( A_{ij} - \frac{k_ik_j}{2m} \right) \delta(c_i,c_j),
    \label{eqn:modularity}
\end{eqnarray}
where $c_i$ is the community of which node $i$ is a member, $k_i$ is the degree of node $i$, $\mathbf{A}=[A_{ij}]$ is the adjacency matrix\footnote{The elements of the adjacency matrix, $A_{ij}$, are equal to $1$ if an edge exist between nodes $i$ and $j$, and $0$ otherwise.} $m$ is the number of edges, $n$ is the number of nodes, and $\delta$ is the Kronecker delta function\footnote{The Kronecker delta function, $\delta(\cdot)$, takes the value $1$ if  the elements are equal and $0$ otherwise.}. 

Modularity is often used to compare different community structures in the same or similar networks. Due to the null model used, modularity can not be used to compare the community structures of different networks, as its maximum value is determined by the network structure. Often higher modularity is taken as an indication of a better community structure as it is more different compared to the random null model. As such, the modularity can only be used as a comparative measure and has drawbacks including difficulties in interpretation, that random networks usually have higher maximum modularity than real-world networks, and the previously discussed resolution limit.

It can be shown that modularity optimization, i.e.\ finding the optimal community structure, is an NP-complete problem \cite{citeulike:5919261}. Thus the problem of community detection is time consuming for large networks and good approximations are necessary for detecting communities with reasonable computational effort. In the following section, three different methods are introduced for detecting communities in networks. These methods are examples of heuristic and stochastic methods for relaxing the NP-complete problem. It is also possible to show that the modularity has many local maxima \cite{Good2009} making the identification of a global maximum very difficult.

\subsection{Algorithmic community detection}
As previously discussed, large efforts have been given the problem of automatically identifying communities in networks. We refer to these methods as algorithmic community detection methods, in contrast to earlier manual methods pioneered by Refs.\ \cite{Weiss1955,Rice1927}. A large number of methods have been proposed based on concepts from e.g.\ the fields of computer science, discrete mathematics, statistical physics, and statistics. In this paper, we consider three different methods: the q-Potts based spin glass algorithm (SP) introduced in Ref.\ \cite{Reichardt}, the greedy agglomerative method (GA) proposed in Ref.\ \cite{citeulike:2390590}, and the fast stochastic method of propagation of labels (LP) presented in Ref.\ \cite{citeulike:1724653}.

The SP-algorithm is based on a q-Potts spin glass and communities are detected by minimizing the energy of the following Hamiltonian
\begin{equation}
	\mathcal{H} = -J \sum_{i=1}^n \sum_{j=1}^n A_{ij} \delta(c_i,c_j) + \gamma \sum_{k=1}^q { s_k \choose 2},
	\label{eqn:hamiltonian}
\end{equation}
where $J$ and $\gamma$ are \textit{coupling parameters}, $\delta(\cdot)$ is the Kronecker delta function, and $s_k$ is the number of spins in state $k$. The size of the detected communities is determined by the ratio\footnote{In the following simulation experiments, this ratio is set as unity and therefore only communities with sizes larger than $\sqrt{m}$ where $m$ is the number of edges can be detected. \cite{Reichardt}} between the two coupling parameters. This is due to the fact that the first term with coupling factor $J$ favors many edges inside communities and few between communities, The second term which is scaled by $\gamma$ favors a uniform distribution of nodes in communities. \cite{Reichardt}

The configuration of spins (communities) that minimizes the Hamiltonian in (\ref{eqn:hamiltonian}) is found using simulated annealing \cite{annealing}. The system is initialized at the temperature $T_0=1$ and cooled using the cooling factor $0.99$ until the final temperature $T_t=0.1$ is obtained. As simulated annealing is quite computer intensive, this algorithm has a high complexity of at least $\mathcal{O}(n^{2+\theta})$ with $\theta=1.2$ on a sparse network. The advantage of this method is that it is known to often find good approximations of the global minimum of the Hamiltonian and therefore also good approximations of the community structure. 

The GA-method greedily merges pairs of nodes/clusters using agglomerative hierarchical clustering. The order in which nodes are merged is governed by the modularity measure, which is calculated for all  possible merges and the resulting merge is determined by the pair that yield the highest increase. This greedy method is quite computation intensive as many possible merges must be evaluated and it is not certain that the optimal solution yielding the maximum modularity is found. The complexity for this algorithm is estimated to be  $\mathcal{O} (n log^2 n)$, which is quite low in comparison with the SP-algorithm. \cite{citeulike:2390590}

\begin{figure}[!h]
	\centering
	\includegraphics[width=0.40\textwidth,angle=0]{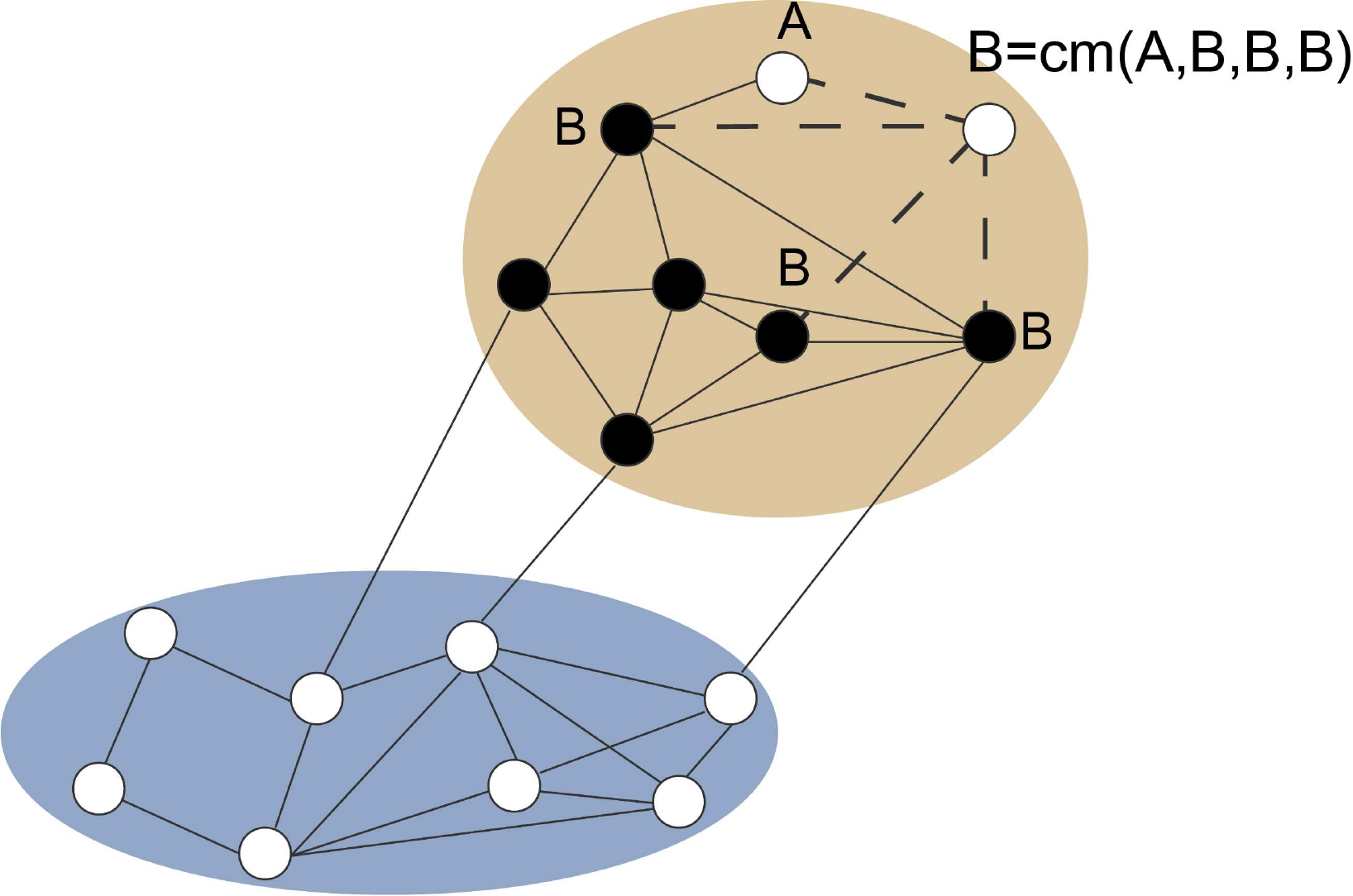}
    \caption{\small{A simple situation in the label propagation algorithm. A node is voted to change its label to B instead of A, as this label is in the majority of the neighboring nodes. cm is a function returning the most common label breaking ties randomly.}}
    \label{fig:labelprop}
\end{figure}

The LP-algorithm is an example of a stochastic method for detecting communities in networks. The method uses labels for each node to decide to which community the specific node belongs. LP is an iterative algorithms that initializes by assigning each node an unique label. The iterative step begins by selecting a node at random and then assigning it a new label using voting (breaking ties randomly) by the labels of the neighboring nodes. The iterative procedure is repeated until no node changes label and thereby an equilibrium is obtained. As communities are groups of nodes more densely connected than with other communities, the labels should propagate and spread in each community. \cite{citeulike:1724653}

The stochastic part of this algorithm is two-fold: firstly the order in which nodes are selected and secondly the random breaking of ties. These two factors are responsible for that the algorithm produces random outcomes. The advantage of this method is that it is very fast, $\mathcal{O}(m)$, where $m$ is the number of edges which is $n \leq m \leq n^2$. Several runs of the LP-algorithm are often combined to counter the stochastic nature of the method. This combination is the essential topic of this paper and it is further discussed in the next section. \cite{citeulike:1724653}

\section{Ensemble methods}
\label{sec:ensemble}
Community detection is a form of clustering of network data, in which nodes are similar by sharing many common neighbors. Clustering in turn is a form of classification, which is extensively used in machine learning and other related areas. In this section, we discuss the important concept of \textit{boosting} used in classification to combine several weaker classifiers into a better classifier using (weighted) voting schemes. Another important concept in classification is \emph{bagging} in which a large number of bootstrap replicated are aggregated to form a robust average. This method has been successfully applied using network data in Ref.\ \cite{citeulike:6597911} and it is therefore likely that boosting will also be applicable to network data.

Boosting has previously been used on clustering methods in e.g.\ bioinformatics to improve the result of clustering analysis. As clustering is similar to community detection, it is fruitful to discuss the \emph{ensemble clustering methods} previously used in data clustering and generalize these for network data. This is the aim of this section which also contains a proposed method for combining several runs of a stochastic community detection algorithm, as the LP method previously discussed, or the structures found by different community detection methods and by bootstrap re-sampled networks.

\subsection{Boosting classifiers}
The idea behind boosting classifiers is to arrange a large number of simple (weak) classifiers into an ensemble (committee), which by a wisdom-of-crowd-effect creates a better classifier. In many cases, the resulting classifier performs much better than a simple more complex classifier. This makes boosting a powerful, yet simple method to greatly improve the classification accuracy.

Another related method to boosting is ensemble learning, in which a number of different weak classifiers are combined into an ensemble without any re-sampling or re-weighting. This family of methods suits our setting better and is the basis for the following discussion on ensemble clustering. It is however important to put some effort into trying to explain why a group of simpler classifiers may perform much better than single more advanced classifier. Much effort has been devoted to answer this question and some answers have been found for independent classifiers.

The ensemble method is discussed in Ref.\ \cite{citeulike:816658} for use with neural networks, which are trained using some data set. It is possible to show that the training problem is an optimization problem with many local minimum (as in the case with modularity maximization). Therefore the weightings found can differ largely for solutions with almost the same error rate. By combining many of these weightings, the authors could show large improvements in overall accuracy. Assume that each classifier has some probability of classification error, $p$, therefore the probability of finding exactly $k$ classification errors in $N$ classifiers is given by
\begin{equation}
	{N \choose k} p^k \left( 1 - p \right)^{N-k},
\end{equation}
and by applying a simple majority voting rule, the corresponding probability of $k$ mis-classifications in an ensemble with $N$ classifiers is
\begin{equation}
	\sum_{k > N/2}^N {N \choose k} p^k \left( 1 - p \right)^{N-k}.
\end{equation}
It is further stated in Ref.\ \cite{citeulike:816658}, that it is possible by induction to prove that this probability is decreasing with increasing $N$ when $p < \frac{1}{2}$. This means that when each classifier is better than a random classification and independent of other classifiers, an arbitrarily error rate can be selected by varying the number of classifiers used in the ensemble. The assumption of independence is seldom valid in practical applications but the method still works when each classifier perform better than chance.

The error rate of an ensemble is further discussed and calculated for dependent classifiers in Ref.\ \cite{Sollich1996}, where the \emph{ensemble generalization error}, $\epsilon$, is expressed as
\begin{equation}
	\epsilon = \sum_k w_k \left[ \left( y(x)-f_k(x) \right)^2 - \left( f_k(x) - \bar{f}(x) \right)^2 \right], 
	\label{eqn:ensembleerror}
\end{equation}
where $y(x)$ is the label of observation $x$, $f_k(x)$ is the label given by classifier $k$ and $\bar{f}(x)$ is the weighted ensemble average,
\begin{equation}
	\bar{f} = \sum_k w_k f_k(x),
\end{equation}
for some weights $w_k$ for classifiers $k=1,2,\ldots,n$. 

The first term in Eq.\ (\ref{eqn:ensembleerror}) is the (weighted) average of the generalization errors of the individual predictors and the second is the weighted average of ambiguities. The latter contains all correlations between the different classifiers. Finally the relation shows that the more predictors differ, the better is the performance of the ensemble. This explains why an ensemble of classifiers performs better than a more advanced single classifiers, as the error rate can be decreased by increasing the number of classifiers included in the ensemble. \cite{Sollich1996}

\subsection{Ensemble clustering}
As classification is related to clustering, it is reasonable that these ensemble methods are useful in clustering as well. In ensemble clustering, the problem is often to combine an ensemble of clusterings generated by e.g.\ some re-sampling method (bootstrap) \cite{citeulike:1632077}. The combination should return the average or aggregated properties of the clusterings found in the ensemble. A method for finding ensemble clusterings is proposed by Ref.\ \cite{strehl03} called \textit{Instance-based Ensemble Clustering} (IBEC). Other important examples of ensemble clustering methods are found in Refs.\ \cite{citeulike:1319014}, but are not used in this paper.

\begin{definition}[IBEC]
Given an ensemble of clusters, $\mathbf{x}=\{\mathbf{x}^{(1)},\ldots,\mathbf{x}^{(r)}\}$, IBEC constructs a fully connected (complete) graph, $G=(V,\mathbf{F})$, where $V$ is a set of $n$ nodes and $\mathbf{F}=[F_{ij}]$ is a \textit{frequency matrix} with $F_{ij}$ as the frequency of instances that nodes $i$ and $j$ are placed in the same cluster.
\label{def:IBEC}
\end{definition}

The IBEC method aggregates the clusterings by constructing a graph where each node represents a data point and each edge indicates that the two connected nodes have been clustered together. The frequency with which the two nodes have been clustered together acts as a weight or similarity for the resulting edge, see \textit{Definition \ref{def:IBEC}} for details. The nodes are finally partitioned into clusters using agglomerative hierarchical clustering with some linkage rule, or by a graph partitioning method as the Kernighan-Lin algorithm \cite{newman2010}.

\subsection{Node-based Fusion of Communities}
\label{sec:algorithm}
In this paper, we propose a generalization of IBEC for network data and for fusing different community structures (subgraphs) into a final representation. This final community structure should indicate the most probable structure as it is the aggregated information from many candidates. \textit{Node-based Fusion of Communities} (NFC) is similar to the previously discussed IBEC but use a special linkage rule to account for the special nature of network data, i.e.\ nodes can not  be placed in the same community if they are not connected by a sufficiently short path.

\begin{figure}[h]
	\centering
	\includegraphics[width=0.40\textwidth,angle=0]{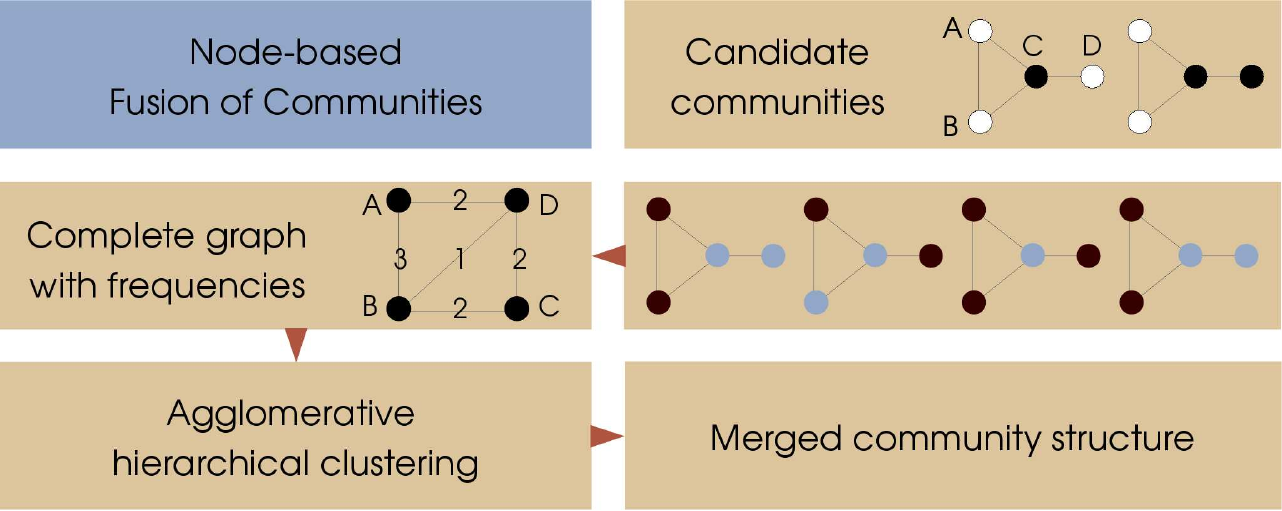}
	\caption{\small{Node-based Fusion using agglomerative hierarchical clustering.}}
    	\label{fig:ibcm}
\end{figure}

The NFC-method is outlined in \textit{Figure \ref{fig:ibcm}}. Firstly, a complete graph, $G=(V,\mathbf{F})$, is constructed using the data from candidate communities, which are the output from some community detection algorithm(s). The set of nodes, $V$, is the original set of nodes in the network, and the set of edges $\mathbf{F}$ now indicate that two nodes have been found in the same community. The matrix, $\mathbf{F}=[F_{ij}]$, where the element in row $i$ and column $j$, $F_{ij}$ is the frequency of the event that nodes $i$ and $j$ has been found in the same candidate community. 

This new graph is clustered using agglomerative hierarchical clustering using a special linkage rule. This is necessary as recalculating is needed for determining the frequency of that the nodes have been placed in the same community as the meta-cluster, i.e.\ a cluster of some merged nodes.

The frequency between the merged nodes (cluster) $l$ and the other nodes or clusters, $v_1,v_2,\ldots,v_{n_l}$, is found by
\begin{equation}
	F_{k,l} = \left|\bigcap_{k} M_{kl} \right|,
\end{equation}
where the \textit{membership matrix}, $\mathbf{M} = [M_{ik}]$, where $M_{ik}$ is the community in which node $i$ is a member in candidate network $k = \{j,i_1,\ldots,i_{n_l}\}$. That is, $F_{k,l}$ is the number of occurrences where all nodes (in both clusters) are in the same candidate cluster. Using this linkage rule incur some information loss as information of individual nodes is lost in the meta-cluster. 

The result of the hierarchical clustering algorithm is a dendrogram and a list of merges. The clustering corresponding to the maximum modularity is taken as the communities found in the merged candidate networks. The complexity of NFC is determined by the hierarchical clustering algorithm and therefore is $\mathcal{O}(n^2)$.

\subsection{Estimating certainty in community structures}
By using the output from the NFC-method, it is possible to estimate the certainty of the hypothesis that a node belongs to a certain community. This is especially important for nodes lying in the borderland between communities. Perhaps it is equally likely that the node belongs to another neighboring community. It is also important in structures similar to chains and tree in the network. These nodes are naturally quite sensitive to uncertain edges because they have few neighbors. Nodes having an uncertain community membership can be found using the candidate communities. If a node is found quite often in two different communities, the confidence that it has been classified correctly is low as it is very sensitive to the network structure.

For the \textit{node-based method}, the nodes are merged in a hierarchical manner, the two first nodes are the most similar and for each merge the nodes get more dissimilar. If a node was merged early into a community, it is less likely that it would belong to another community. Therefore a qualitative measure of the certainty that a node $i$ belongs to a community is found as $b_i=t_i^{-1}$ where $t_i$ is the number of merges needed before the node $i$ is added to the community. A larger value of this score indicates an early merge and therefore a more certain merge.

\section{Simulation experiments}
\label{sec:experiments}
This section contains details regarding the conducted simulation experiments using the proposed method for merging community structures, the NFC method. We propose to use this method in combination with the stochastic LP-algorithm and call this combination Label Propagation-Node-based Fusion of Communities (LP-NFC). This section discuss some preliminary methods including the generation of synthetic networks and performance measures. Comparative studies of the proposed LP-NFC algorithm with the SP and GA-algorithms are presented and the community detection methods are evaluated by performance and computational complexity. 

\subsection{Synthetic networks with community structure}
\label{sec:syntheticnetworks}
The proposed method is demonstrated using synthetic (random network with community structures) networks with a community structure. The \textit{synthetic network} model used in this paper is adopted from Refs.\ \cite{citeulike:5333047,citeulike:3454707}. The authors have constructed algorithms to generate artificial networks with community structures, which has become a standard benchmark for community detection using synthetic networks.  The networks are generated using six different input parameters, shown in \textit{Table \ref{tbl:randomnet}}, together with the values used in the following experiments. These parameters allow for the generation of families of networks with desired properties.

\begin{table}[!h]
\begin{center}
\begin{small}
  \begin{tabular}{ccl}
  \hline
	Variable & Value & Description \\
  \hline
$n$ &-& number of nodes in the network \\
$\bar{k}$ & 15 & mean degree of each node \\
$k_{\max}$ & 50 & maximum degree  \\
$\mu$ &- & mixing parameter \\
$c_{\min}$ & 20 & minimum size of a community \\
$c_{\max}$ & 50 & maximum size of a community \\
$\beta$ & 1 & exponent of community size distribution \\ && (typically $1 \leq \beta \leq 2$ in real-world networks) \\
$\gamma$ & 2 & exponent of degree distribution \\ && (typically $2 \leq \gamma \leq 3$ in real-world networks) \\
  \hline
  \end{tabular}
  \caption{\small{The parameters used for generating synthetic networks in the simulation studies using the algorithm from Ref.\ \cite{citeulike:5333047}.  These parameters create networks similar to Newman-Girvan benchmarks. The mixing parameter, $\mu$, and number of nodes $n$ are varied during the experiments. \cite{citeulike:1043460} }}
  \label{tbl:randomnet}
\end{small}
\end{center}
\end{table}

The mixing parameter, $\mu$, is the fraction of edges between the different communities and $1-\mu$ is the fraction of intra-community edges. A small mixing parameter corresponds to well-separated communities, the extreme is when $\mu=0$ and only disjoint communities exist. As $\mu$ increases, the communities become more difficult to detect, until $\mu=0.5$ when no communities exist in the network according to the adopted definition of a community in \textit{Definition \ref{def:community}}.

The algorithm to generate the synthetic networks consists of five different steps. A simplified version, see Ref.\ \cite{citeulike:5333047} for the full version, is as follows: Firstly, generate the degree of each node by sampling from a power-law distribution with parameter, $\gamma$, satisfying, $\bar{k}$ and $k_{\max}$. Secondly, generate the size of each community from a power-law distribution with parameter, $\beta$, such that all nodes are members of a community and the community sizes are consistent with the parameters, $c_{\min}$ and $c_{\max}$.
\begin{enumerate}
	\item using the configuration model, assign edges between all nodes such that the degree of all nodes are satisfied,
	\item randomly distribute the nodes to the communities in the network,
	\item rewire the edges between nodes until the mixing parameter, $\mu$ is satisfied.
\end{enumerate}

The drawback of this algorithm is the lack of triangles observed in real-world social networks, which result in a sparser network than in empirically found networks. The advantage is that synthetic networks enable the study of how the mixing parameter is correlated with the effectiveness in finding communities in uncertain networks. The algorithm has a linear complexity, $\mathcal{O}(n)$, and can therefore be used to simulate large networks with community structures that are consistent with real-world social networks. \cite{citeulike:5333047}

\subsection{Comparing community structures}
Supervised measures are used to evaluate the performance of the community detections methods used in the simulation experiments. This is possible due to that the \emph{correct} community structured is returned by the synthetic networks algorithm, thus external information is available for supervised methods. Traditionally, supervised methods have included precision and recall which are commonly used in classification and clustering to evaluate methods and algorithms. The drawback with these traditional methods are that it is difficult to find the matching pair of labels from the obtained solution and the externally provided labels. For example, a single community detected by the community detection algorithms may correspond to two different labels in the external information. The problem is to select the external label to match this obtained community with. Previously the corresponding label has been taken as the set with the largest overlap, therefore only obtaining approximate performance measures.

In this paper, the \textit{Normalized Mutual Information} (NMI) is instead used to measure the performance of the different community detection algorithm. An additional method using correlation is proposed, as it is a simpler measure and later discussed also yields result similar to the former measure.

\subsubsection{Normalized Mutual Information}
The NMI-measure originates in information theory and can be interpreted as how much is known about the external labeling given the obtained solution and vice versa. We follow Ref.\ \cite{citeulike:6252578} to defined the NMI-measure.

Let $\hat{\mathbb{I}}(\mathcal{C},\mathcal{L})$ denote the NMI-measure where $\mathcal{C}$ is the obtained community structure and $\mathcal{L}$ is the external labeling. Assume that $\mathcal{L}=\{ l_{i} \}$ where $l_i$ is the label of node $i$ and the same for $\mathcal{C}$ with the obtained community membership of node $i$. Further assume that $l$ and $c$ are the realizations of some random variables, $L$ and $C$, with some (joint) probability distributions as
\begin{eqnarray}
	\mathbb{P}(l,c)= \mathbb{P}(L=l,C=c) &=& \frac{|\mathcal{L} \cap \mathcal{C}|}{n}, \\
	\mathbb{P}(l) = \mathbb{P}(L=l) &=& \frac{|\{ l_i \in \mathcal{L} : l_i = l \} |}{n}, \\
	\mathbb{P}(c) = \mathbb{P}(C=c) &=& \frac{| \{ c_i \in \mathcal{C} : c_i = c \}| }{n},
\end{eqnarray}
where $|\{ l_i \in \mathcal{L} : l_i = l \}|$ is the number of elements in $\mathcal{L}$ which equals the label $l$ with the corresponding for $c$, and $n$ is the total number of nodes, $n=|\mathcal{L}| = |\mathcal{C}|$. The \textit{mutual information}, $\mathbb{I}(C,L)$, is defined by

\begin{equation}
	\mathbb{I}(C,L) = \sum_l \sum_c \mathbb{P}(c,l) \log \left( \frac{ \mathbb{P}(c,l)}{\mathbb{P}(c) \mathbb{P}(l)} \right),
	\label{eqn:mutualinfo}
\end{equation}
where the sums are taken over all assumed values of $l$ and $c$, and $\log(\cdot)$ is the logarithm (with base $2$). The NMI-measure, $\hat{\mathbb{I}}(\mathcal{C},\mathcal{L})$, between the obtained community structure, $\mathcal{C}$, and the externally given labels, $\mathcal{L}$, is
\begin{equation}
	\hat{\mathbb{I}}(\mathcal{C},\mathcal{L}) = 2 \frac{ \mathbb{I}(\mathcal{C},\mathcal{L})}{\mathbb{H}(\mathcal{C}) + \mathbb{H}(\mathcal{L})},
\end{equation}
which equals zero if the community structures are independent and unity if they are equivalent. The \textit{entropy}, $\mathbb{H}(X)$, of a random variable $X$ defined as
\begin{equation}
	\mathbb{H}(X) = - \sum_x \mathbb{P}(x) \log \left( \mathbb{P} (x) \right).
	\label{eqn:entropy}
\end{equation}

\subsubsection{Correlation}
The correlation is used to calculate a measure of how the rows in the matrices tend to be similar \cite{jain88}. This type of measure has previously been used in comparing clustering and classification methods. Let $\mathbf{N}=[N_{ij}]$ denote the \emph{neighborhood matrix} where $N_{ij}=1$ if nodes $i$ and $j$ are found in the same cluster and $0$ otherwise. The mean correlation $\bar{\rho}(N,\hat{N})$, between the two matrices, $N$ and $\hat{N}$, is found as the mean of the \textit{Pearson correlations}, $\rho_i$, for each row

\begin{equation}
	\bar{\rho}(N,\hat{N}) = \frac{1}{n} \sum_{i=1}^n \rho_i(N_i,\hat{N_i}) =\frac{1}{n} \sum_{i=1}^n  \frac{\mathsf{Cov} (N_i, \hat{N}_i)}{\mathbb{V}[N_i] \mathbb{V}[\hat{N}_i]},
\end{equation}

where $\mathbb{V}(\cdot)$ denotes the variance. Using the covariance between each element in each row and the expected value (mean) of the that specific row

\begin{equation}
	\mathsf{Cov} (N_i, \hat{N}_i) = \frac{1}{n} \sum_{j=1}^n (N_{ij} - \mathbb{E}[N_i]) (\hat{N}_{ij} -  \mathbb{E}[\hat{N}_i]),
\end{equation}
where $\mathbb{E}(\cdot)$ denotes the expected value. By letting $\mathbf{N}$ be the \textit{ideal neighborhood matrix}, $\mathbb{N^*}=[N^*_{ij}]$, with $N^*_{ij}=1$ when $l_i=l_j$ (equal external labels) but zero otherwise and $N^*_{ii}=0$. Finally letting $\hat{\mathbf{N}}$ denote the neighborhood matrix from the obtain solution, results in another supervised method for comparing the performance of community detection methods. As the NMI, this measure scales to unity when a perfect match is found and if the measure assumes the value zero then no matches are found.

Both the NMI and correlation measures have the advantage over methods like precision and recall, that no identification/linkage of labels are needed. This therefore removes the need of using approximate methods as largest overlap to identify which label that is external given best matches the obatined labels from the community detection method.

\subsection{Convergence properties}
A first important question to answer is how many runs of the LP algorithm need to be merged by the NFC method to obtain stable solutions. This is the question which is to be answered in this section, before any performance comparisons can be made.

\begin{figure}[t]
	\includegraphics[width=0.40\textwidth]{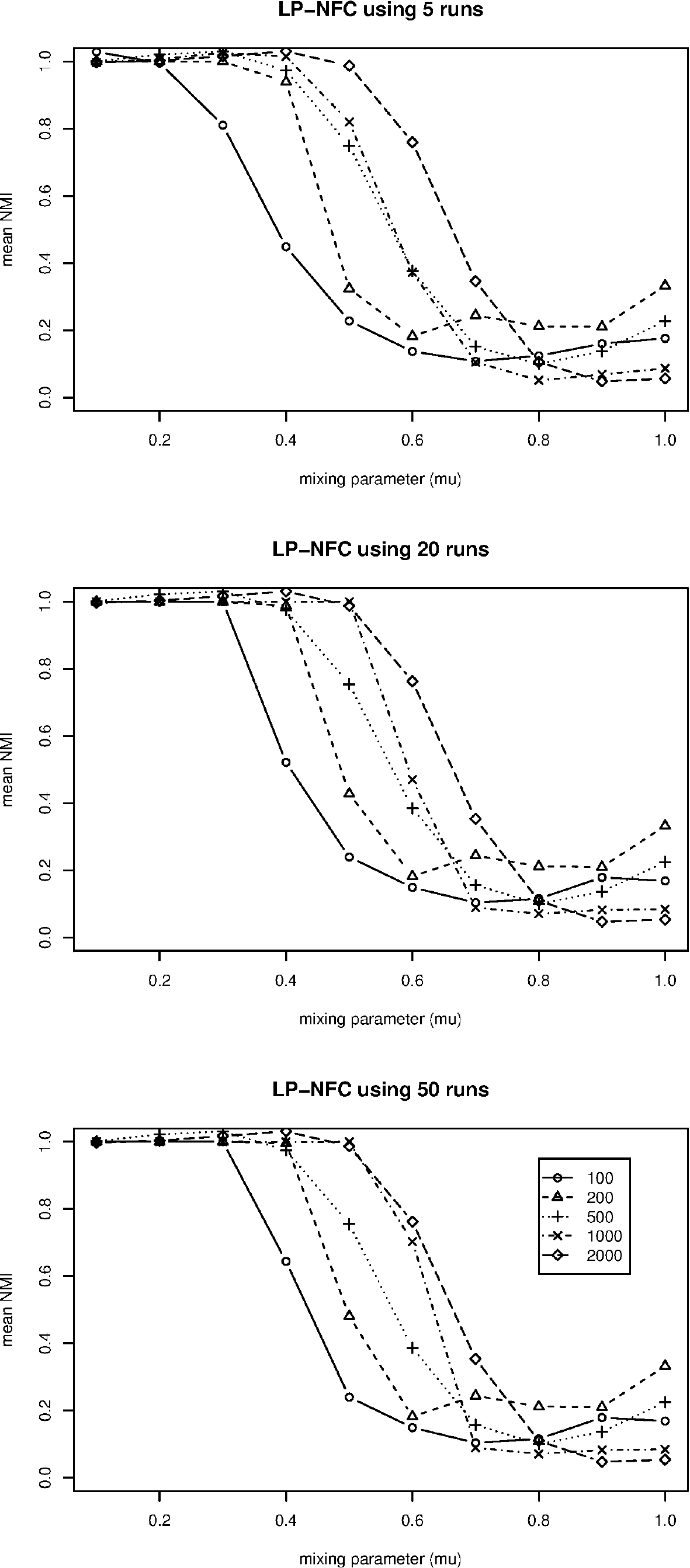}
	\caption{The NMI for the LP-NFC method with three different number of samplings used on synthetic networks with community structures. The graphs are found as an average of $100$ runs on networks with $n \in \{100, 200, 500, 1000, 2000\}$ nodes and mixing parameter, $\mu \in \{0.1,0.2,\ldots,1.0\}$. \label{fig:nosamples}}
\end{figure}

In \textit{Figure \ref{fig:nosamples}}, we present the NMI measures for several different runs of the proposed LP-NFC method. In each run, the number of nodes $n$ is varied between $100$ and $2000$ nodes, the mixing parameter is varied between $0$ and $1$, and finally the number of merged runs $n_r$ is varied between $5$ and $50$.

As the number of samplings increase some of the curves shift rightwards, which indicates better performance in finding the correct structures in network with more diffuse community structures. Remember, that higher mixing parameter indicate more diffuse community structure, that are more difficult to detect. The largest movement is found in the curve corresponding to $n=1000$ nodes. The conclusion is that more samplings are needed in networks with more nodes than in networks with fewer nodes. This as the NMI for the smaller networks are more or less constant with respect to the number of samplings. 

This corresponds to what is known from standard Monte Carlo-methods, that it is possible to decrease the statistical error by increasing the number of samples. This is only possible up to a certain level, before the systematic errors dominates the statistical error. We conclude that $50$ samplings are a good choice due to this result as well as required computational time. 

\subsection{Comparisons with GA and SP}
Continuing with comparing the proposed method to merge (aggregate) a number of runs by the LP-algorithm with the commonly used SP and GA-algorithms. The algorithms have been evaluated using the previously discussed synthetic networks with varying number of nodes, $n \in \{100, 200, 500, 1000, 2000\}$, and mixing parameter, $\mu \in \{0.1,0.2,\ldots,1.0\}$. The obtained community structure is compared with the labeling outputted by the synthetic network algorithm using, the previously discussed. NMI and correlation measures. 

The most important feature in the following figures are the shift from high values of NMI and correlation to lower. The critical value of the mixing parameter, $\mu_c$, clearly depends on the size of the network, $n$, and differ between the compared algorithms. 

\begin{figure}[t]
	\includegraphics[width=0.40\textwidth]{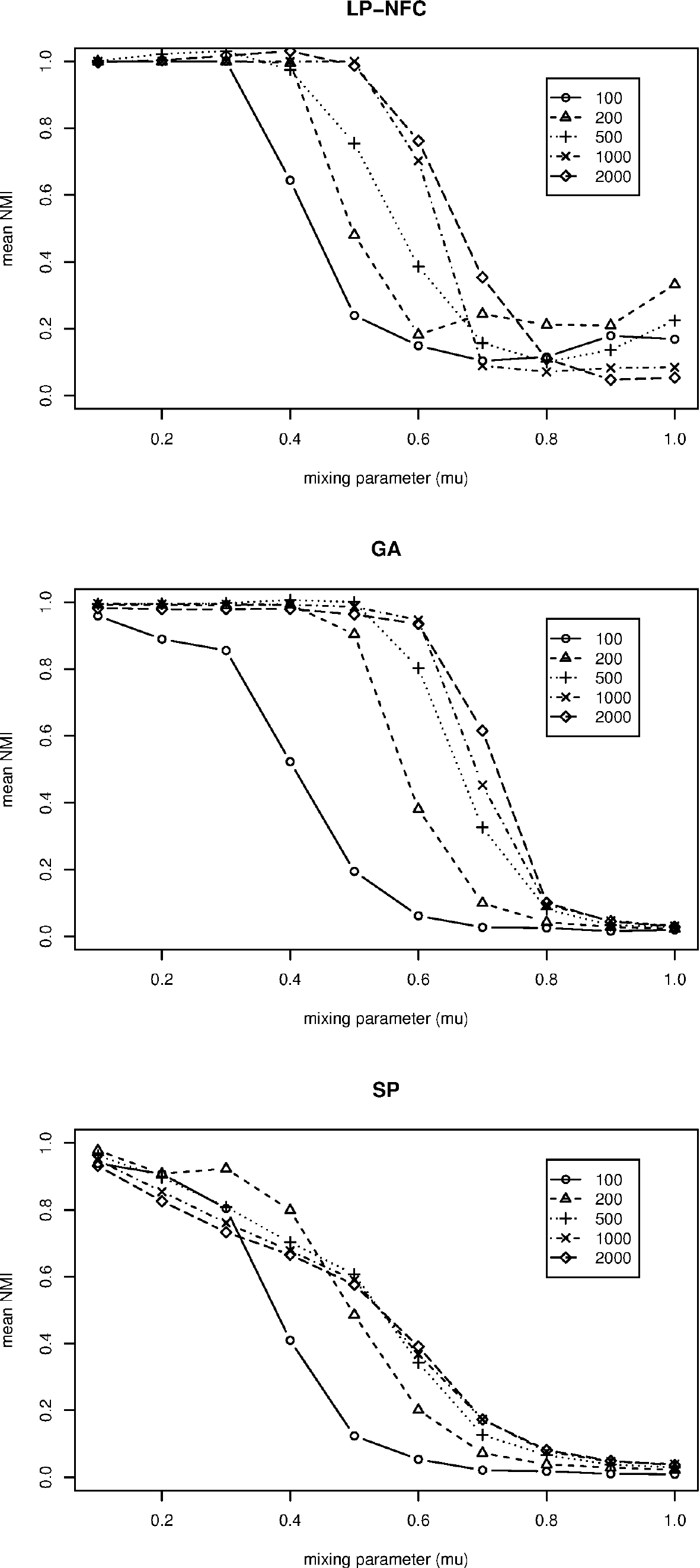}
	\caption{The NMI for three different community detection methods used on synthetic networks with community structures. The graphs are found as an average of $100$ runs on networks with $n \in \{100, 200, 500, 1000, 2000\}$ nodes and mixing parameter, $\mu \in \{0.1,0.2,\ldots,1.0\}$. \label{fig:LPvGAvSP}}
\end{figure}

In \textit{Figure \ref{fig:LPvGAvSP}}, the methods are compared using the mean NMI from $100$ runs at each value of the number of nodes and mixing parameter. The profiles of the LP-NFC and GA-algorithms are quite similar in appearance compared with the SP-algorithm. The latter is previously known to perform worse on synthetic networks than on real-world networks. This is clearly visible by that the NMI values quickly fall of in comparison with the other two algorithms, that seems to have more or less constant NMI until the critical mixing parameter. 

Another feature worth noting is the tail behavior for the LP-NFC-algorithm at high mixing parameters. The NMI value for the other two algorithms quickly drops to zero after the shift in performance, but the LP-NFC algorithm continues to have non-zero NMI. This is particularly visible for the smaller networks with $100$, $200$, and $500$ nodes and is probably some artifact from the stochastic nature of the LP-algorithm.

Comparing the LP-NFC and GA-algorithms, we conclude that the performance is similar between these two methods and are superior to the SP-algorithm. Continuing, with another comparison using the correlation and the computational complexity to find the preferred method. 

\begin{figure}[t]
	\includegraphics[width=0.40\textwidth]{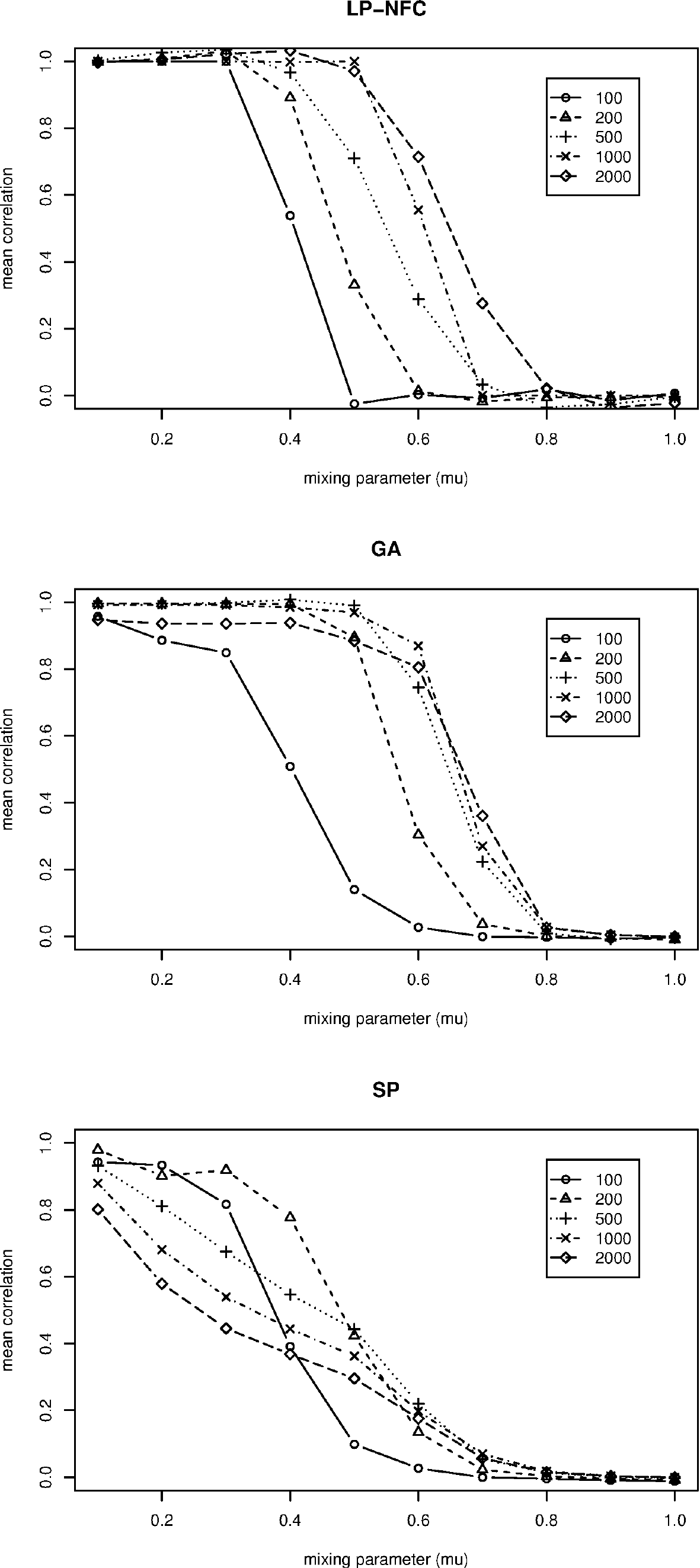}
	\caption{The correlation for three different community detection methods used on synthetic networks with community structures. The graphs are found as an average of $100$ runs on networks with $n \in \{100, 200, 500, 1000, 2000\}$ nodes and mixing parameter, $\mu \in \{0.1,0.2,\ldots,1.0\}$. \label{fig:LPvGAvSPcor}}
\end{figure}

In \textit{Figure \ref{fig:LPvGAvSPcor}}, methods are compared using the mean correlation from $100$ runs at each possible number of nodes and mixing parameter. This figure is quite similar to the previous but with some differences in the LP-NFC algorithm, where the mean correlation drops to zero (as for the other two algorithms). The artifacts at high mixing parameters have therefore vanished and seems to be to be related with the use of the NMI-measure. 

Most of the analysis from the NMI-measure remains with the correlation measure as well. It is perhaps even more apparent that the GA-algorithm is less sensitive to the number of nodes than the LP-NFC-algorithm. This is seen by the densely packed correlation curves in the GA-algorithm that is not visible in the LP-NFC-algorithm. 

The SP-algorithm continues to under-perform in comparison with the two other methods. As previously discussed, this is probably the result of using the synthetic networks, as the method perform well for real-world networks. These two types of networks differ in some important aspects, as for example the number of triangles which could explain the poor performance of the SP-algorithm. 

\begin{figure}
	\includegraphics[width=0.40\textwidth]{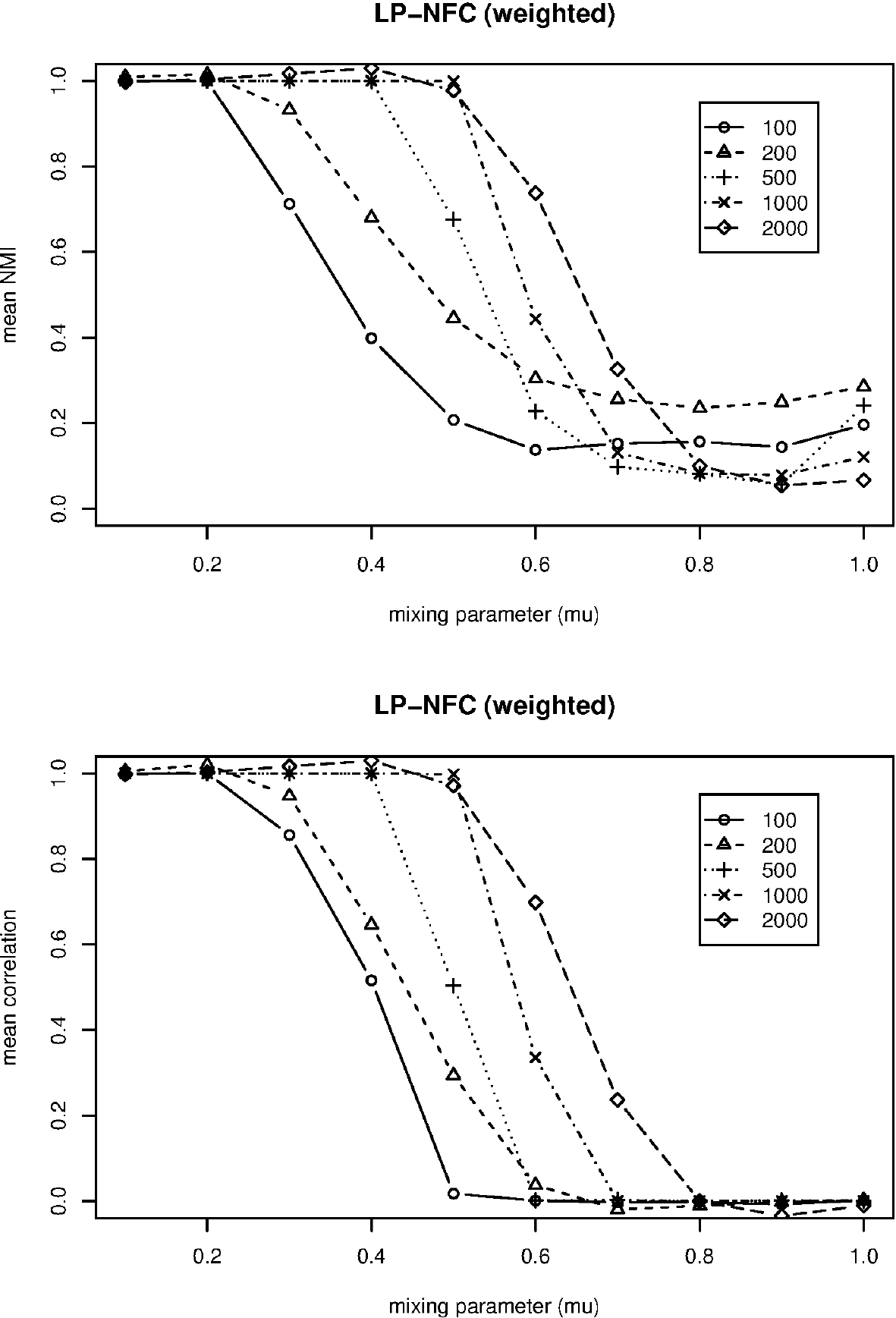}
	\caption{The NMI of the modularity-weighted LP-NFC method. The graph are found as an average of $100$ runs on networks with $n \in \{100, 200, 500, 1000, 2000\}$ nodes and mixing parameter, $\mu \in \{0.1,0.2,\ldots,1.0\}$. \label{fig:weighted}}
\end{figure}

Concluding this analysis, we suggest an improvement to the LP-NFC-algorithm by weighting each frequency by the normalized modularity. For each candidate community structure, the modularity is calculated and normalized with the maximal value of the modularity. This generates a set of weights, $w_i \in [0,1]$, for each candidate community structure and the elements in the frequency matrix is weighted by $w_i$. This gives more weight to nodes that have been placed in the same community in structures with higher modularity than in situations with lower modularity. As the modularity indicates the quality of the community structure found, this could generate better results. 

This modification has been evaluated in the same manner as the previous version and the results from evaluation by the NMI and correlation measures are shown in \textit{Figure \ref{fig:weighted}}.

\subsection{Time complexity}
Important aspects of community detection algorithms are performance and computational complexity. A good algorithm should have a low computational complexity, which is equivalent to scalability to larger networks. Performance and accuracy are also desirable properties of community detection methods. LP has a low complexity and performance, it is therefore interesting to determine the complexity of the LP-NFC. This especially as the LP-NFC method has performed well in comparison with SP and GA-algorithms.

It is previously known that the GA-algorithm has a rather low computational complexity, $\mathcal{O}(n log^2 n)$, for most networks. This is especially true with the modifications described in Ref.\ \cite{citeulike:1919636}. The simulation experiments in this paper are done in the software R using the implementations offered by the \texttt{igraph}-package, which are not optimized for low computational complexity. The following comparison is therefore just preliminary and a better implementation of LP-NFC is needed for making better comparisons. 

The LP-NFC-algorithm is based on two different steps, the first is $p$ runs of the LP-algorithm which then is merged using agglomerative hierarchical clustering. Using the linkage rule considered in this paper, the complexity of the latter algorithms is $\mathcal{O}(n^2 \log n)$ or $\mathcal{O}(n^2)$. The total complexity of the method is then, $\mathcal{O}(pm+n^2 \log n)$, where $m$ is the number of edges with $m < n^2$ and $p$ is some suitable number of merged runs e.g.\ $p=n$. This gives a theoretical computational complexity of approximately $\mathcal{O}(n^2 \log(n))$. 

\begin{figure}
	\includegraphics[width=0.40\textwidth]{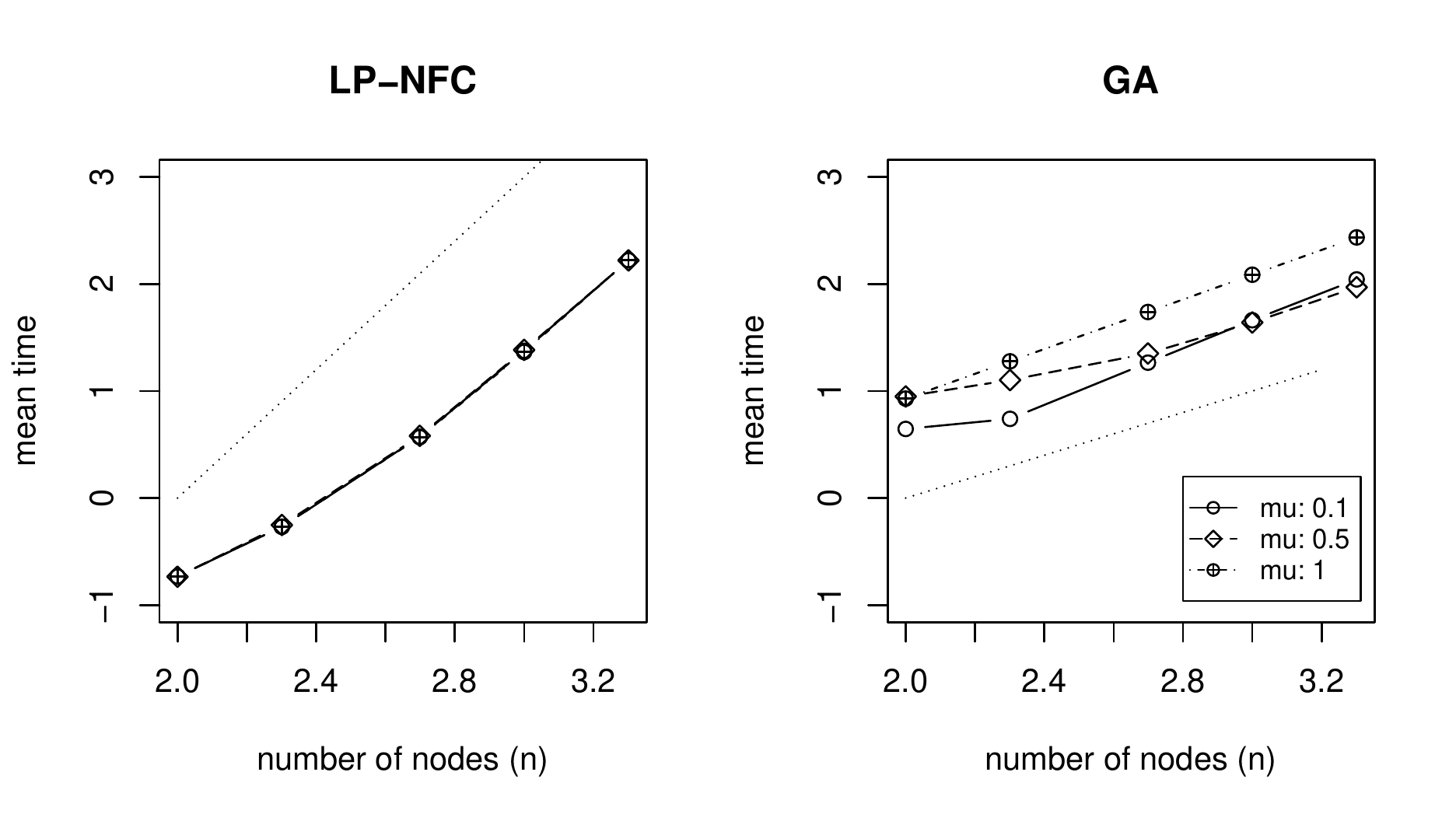}
	\caption{The running time of the LP-NFC and GA-algorithms. The graph are found as an average of $30$ runs on networks with $n \in \{100, 200, 500, 1000, 2000\}$ nodes and mixing parameter, $\mu \in \{0.1,0.2,\ldots,1.0\}$. The dotted lines are reference curves for $\mathcal{O}(n^3)$ for the LP-NFC algorithm and $\mathcal{O}(n)$ for the GA-algorithm. \label{fig:computationaltime}}
\end{figure}

The running time of the LP-NFC and GA-algorithms is shown in \textit{Figure \ref{fig:computationaltime}} for different numbers of nodes, $n$, and mixing parameters, $\mu$. The LP-NFC-algorithm have a rather high complexity in its current implementation as previously discussed. It is approximately $\mathcal{O}(n^3)$, which is higher than the theoretical value. The GA-algorithm has about linear computational complexity, as previously discussed by Ref.\ \cite{citeulike:1919636}. Some other interesting aspects is that the LP-NFC-algorithm is a lot faster for smaller networks (have a smaller constant term than the GA-algorithm) and the impact of the mixing parameter. In the GA-algorithm the mixing parameter has a rather high influence on the running time of the community detection method. This effect is not visible for the proposed method in this paper.

The SP-algorithm is neglected in this comparison because of implementation differences, making this algorithm a lot faster than the other two. This making the comparison difficult but the theoretical complexity of the SP-algorithm is approximately $\mathcal{O}(n^{3.2})$ for sparse networks where $m << n^2$. 

We conclude by that the computational complexity of the current implementation of LP-NFC can be improved but the theoretical limit is still higher than for the GA-algorithm. The advantage of the LP-NFC-algorithm is that the running time is not an increasing function of the mixing parameter and is faster for networks smaller than $1000$ nodes.

\section{Concluding remarks}
\label{sec:conclusions}
In this paper, we have presented a method for combining community structures detected in networks, named Node-based Fusion of Communities (NFC). This method has applications including combining several different community detection method, for enhancing the performance of stochastic methods, and for merging communities detected at different scales. The method has been used in combination with the Label Propagation (LP) algorithm and evaluated using simulation studies with synthetic networks.

\begin{acknowledgments}
JD would especially like to thank Sang Hoon Lee, Jari Saramäki, Petter Holme, and Martin Rosvall for helpful discussions, comments, and suggestions during the work underlying this paper. This paper is part of the project \textit{Tools for information management and analysis}, which is funded by the R\&D programme of the Swedish Armed Forces.
\end{acknowledgments}


%

\end{document}